\documentclass{jpp}
\usepackage{graphicx}
\usepackage{epstopdf, epsfig}
\usepackage{hyperref}
\usepackage{color}
\usepackage[normalem]{ulem}
\usepackage{url}

\usepackage[utf8]{inputenc}
\usepackage[T1]{fontenc}
\usepackage{amsmath}

\newcommand{\pgrad}{\nabla_\perp}
\newcommand{\BB}{\boldsymbol{B}}
\newcommand{\const}{\mathrm{const}}
\newcommand{\etabar}{\overline{\eta}}

\shorttitle{Asymptotic quasisymmetric Grad-Shafranov equation}
\shortauthor{N. Nikulsin, W. Sengupta, R. Jorge and A. Bhattacharjee}

\title{An asymptotic Grad-Shafranov equation for quasisymmetric stellarators}

\author{Nikita Nikulsin\aff{1}
  \corresp{\email{nnikulsin@princeton.edu}},
  Wrick Sengupta\aff{1},
  Rogerio Jorge\aff{2},
 \and Amitava Bhattacharjee\aff{1}}

\affiliation{\aff{1}Department of Astrophysical Sciences, Princeton University, Princeton, New Jersey 08544, USA
\aff{2}Department of Physics, University of Wisconsin-Madison, Madison, Wisconsin 53706, USA}

\begin{document}

\maketitle

\begin{abstract}
A first-order model is derived for quasisymmetric stellarators where the vacuum field due to coils is dominant, but plasma-current-induced terms are not negligible and can contribute to magnetic differential equations, with $\beta$ of the order of the ratio of induced to vacuum fields. Under these assumptions, it is proven that the aspect ratio must be large and a simple expression can be obtained for the lowest-order vacuum field. The first-order correction, which involves both vacuum and current-driven fields, is governed by a Grad-Shafranov equation and the requirement that flux surfaces exist. These two equations are not always consistent, and so this model is generally overconstrained, but special solutions exist that satisfy both equations simultaneously. One family of such solutions are the first-order near-axis solutions. Thus, the first-order near-axis model is a subset of the model presented here. Several other solutions outside the scope of the near-axis model are also found. A case study comparing one such solution to a VMEC-generated solution shows good agreement.
\end{abstract}

\section{Introduction}
Quasisymmetry, along with quasi-isodynamicity, is one of the two main approaches to confining particle orbits inside a stellarator \citep{helander2014theory,boozer1998what}. Quasisymmetry is usually defined as a stellarator configuration where the magnitude of the magnetic field satisfies $|\BB| = B(\psi, M\theta - N\phi)$ for some integers $M,N$, where $(\psi,\theta,\phi)$ is a straight field line coordinate system \citep{helander2014theory}. Thus, the magnetic field magnitude is two-dimensional, but the full vector can still be three-dimensional. Such a device can be modelled using the near-axis expansion \citep{garren1991existence,garren1991magnetic,landreman2018direct}, which consists of Taylor-expanding all quantities of interest in either the distance from the magnetic axis (direct coordinates), or the square root of toroidal flux (inverse coordinates) \citep{jorge2020construction}, and balancing the coefficients in the MHD equilibrium equations at each order in the Taylor expansion, resulting in a system of coupled ordinary differential and algebraic equations. However, the imposition of the quasisymmetry constraint in addition to force balance results in the system being overconstrained at all orders beyond the first in the near-axis expansion \citep{garren1991existence}. At second order, this can be rectified by not imposing quasisymmetry directly, but rather optimizing for it. Unlike traditional optimization, second order near-axis optimization is much faster due to the significantly reduced parameter space \citep{landreman2022mapping}. Beyond second order, there is evidence that the Taylor series of the near-axis model begins to diverge \citep{rodriguez2022quasisymmetry}.

Magnetic shear appears at third order in the near-axis expansion, and thus depends on third-order quantities. While, in some cases, setting the third-order quantities to zero and evaluating the shear based on just lower-order quantities provides a reasonable estimate, in other cases such an approach produces large deviations from the shear calculated in numerical solutions \citep{rodriguez2022quasisymmetry}. This presents a challenge for the modelling of quasisymmetric devices, as shear is an important quantity that affects ballooning stability of equilibria, among other things \citep{connor1978shear}.

A complementary model for quasiaxisymmetric stellarators near axisymmetry that can incorporate shear has been derived in a previous paper \citep{sengupta2024asymptotic}. In this paper, we propose a more general model that is applicable to both quasiaxisymmetric and quasi-helically symmetric stellarators far from axisymmetry. In both cases the difficulties related to shear are avoided by not using Taylor expansions like the near-axis model does. Instead, we perform an asymptotic expansion around a vacuum magnetic field $\nabla\chi$, where $\chi$ is the magnetic scalar potential. The dominance of the vacuum field is a natural assumption, given the design principles of stellarators, which aim to confine the plasma without having a large plasma current \citep{freidberg2014ideal}. Our approach is inspired by Strauss' derivation of reduced MHD \citep{strauss1997reduced}, who also expanded around a vacuum field, with the main difference being that Strauss assumes $p = O(\epsilon^2)$, whereas we allow for a higher $\beta$ with $p = O(\epsilon)$, where $\epsilon$ is a small parameter. We will show that such an ordering requires $\chi$ to have a specific form. In the limit of low $\beta$, our model will reduce to the equilibrium limit of Strauss' equations under the assumption of quasisymmetry with our choice of $\chi$. The derivation will closely follow that of the Freidberg high-$\beta$ stellarator model \citep{freidberg2014ideal}, except that Freidberg expands around a purely toroidal field $B_0\widehat{\phi}$, whereas we allow for a more general zeroth-order field. Once the model is derived, we will present several solutions and a numerical validation of one of those solutions.
 
\section{Derivation}\label{sec:deriv}
To begin, we expand the magnetic field around a vacuum magnetic field $\nabla\chi$, where $\nabla^2\chi = O(\epsilon^2)$:
\begin{equation}
    \BB = \left(1 + \frac{B_1}{B_v}\right)\nabla\chi + \BB_{\perp 1} + O(\epsilon^2).\label{eq:Bfield}
\end{equation}
We will order relative to the $\nabla\chi$ term: $B_v = |\nabla\chi| = O(1)$, $B_1 = O(\epsilon)$ and $\BB_{\perp 1} = O(\epsilon)$, where $\epsilon = \max|\BB - \nabla\chi|/B_v \ll 1$ and $\BB_{\perp 1}\perp\nabla\chi$. In addition, the derivative along the vacuum field must be ordered as $\nabla\chi\cdot\nabla = O(\epsilon)$, whereas $\pgrad = \nabla - B_v^{-1}\nabla\chi\cdot\nabla = O(1)$. The derivative ordering can be justified as follows. The the perpendicular length scale will be $\sim r_0$, whereas the length scale along $\nabla\chi$ will be $\sim 2\pi R_0$, where $r_0$ and $R_0$ are the minor and major radii, respectively. Thus, the ratio of the length scales must be less than $1/2\pi$, and no more than 0.1 for any realistic aspect ratio.

Taking the divergence of \eqref{eq:Bfield}, we have $\nabla\cdot\BB = \nabla\chi\cdot\nabla(B_1/B_v) + \nabla\cdot\BB_{\perp 1} = {\pgrad\cdot\BB_{\perp 1}} + O(\epsilon^2)$. Since $\pgrad$ is two-dimensional, a stream function $A = O(\epsilon)$ can be introduced: $\BB_{\perp 1} = \nabla A\times\nabla\chi$. Note that Strauss uses the symbol $\psi$ for his stream function, but we will use $A$ instead to avoid confusion with the flux surface label.

Following Strauss, we can write the current density as follows:
\begin{equation}
    \boldsymbol{j} = \frac{\BB\times\nabla p}{B^2} + j_\parallel\frac{\BB}{B} = \frac{\nabla\chi\times\nabla p}{B_v^2} - \frac{\BB}{\mu_0}\Delta^* A + O(\epsilon^2),\label{eq:curr1}
\end{equation}
where $\Delta^* = B_v^{-2}\nabla\cdot(B_v^2\pgrad)$, as defined in \cite{strauss1997reduced}, and the expression for $j_\parallel$ is easily obtained by dotting the curl of \eqref{eq:Bfield} with $\nabla\chi$ and using the identity $\nabla f\cdot\nabla\times\boldsymbol{U} = -\nabla\cdot(\nabla f\times\boldsymbol{U})$. An alternate expression can be obtained by taking the curl of \eqref{eq:Bfield}:
\begin{equation}
    \boldsymbol{j} = \frac{1}{\mu_0}\nabla\left(\frac{B_1}{B_v}\right)\times\nabla\chi - \frac{\nabla\chi}{\mu_0}\nabla^2 A + \frac{1}{\mu_0}(\nabla\chi\cdot\nabla)\nabla A - \frac{1}{\mu_0}(\nabla A\cdot\nabla)\nabla\chi.
\end{equation}
The last two terms are both $O(\epsilon^2)$. This becomes more obvious for the last term if we write it as $(\nabla A\cdot\nabla)\nabla\chi = \nabla(\nabla\chi\cdot\nabla A) - (\nabla\chi\cdot\nabla)\nabla A$. Equating the perpendicular components of the current at lowest order, we obtain
\begin{equation}
    \left[\nabla\left(\frac{B_1}{\mu_0 B_v}\right) + \frac{1}{B_v^2}\nabla p\right]\times\nabla\chi = 0.\label{eq:perp_fb}
\end{equation}
Unless the second term is a gradient, both terms are linearly independent and must be individually zero at order $\epsilon$ (see Appendix \ref{sec:perp_fb} for details). If the second term is a gradient, we must have either $B_v = B_v(p)$ or $B_v = B_0 + O(\epsilon)$ where $B_0 = \const$; however, the former cannot be true as $|\BB|$ can only be a flux function in axisymmetry \citep{schief2003nested}, and since $|\BB| = B_v + O(\epsilon^2)$, $B_v$ cannot be a flux function either. Thus, in order to satisfy the equation, we must either ensure that both terms are individually zero, i.e. $p = O(\epsilon^2)$ and $B_1 = 0$, or let $B_v = B_0 + O(\epsilon)$. The former corresponds to the equilibrium limit of the Strauss equations, and so we focus on the latter. The equation then yields:
\begin{equation}
    \frac{B_0 B_1}{\mu_0} + p = 0.\label{eq:theta_fb}
\end{equation}
This is simply the lowest-order radial pressure balance condition; similar expressions appear in most reduced MHD models that assume $p = O(\epsilon)$ \citep{freidberg2014ideal,zocco2021magnetic,kruger1998generalized}. Taking the divergence of equation \eqref{eq:curr1} then gives the generalized version of Strauss' equation (26) in the equilibrium limit:
\begin{equation}
    \BB\cdot\nabla\left(\frac{j_\parallel}{B}\right) + \nabla\cdot\left(\frac{\BB\times\nabla p}{B^2}\right) = \frac{-1}{\mu_0}\BB\cdot\nabla\Delta^* A + \nabla\left(\frac{1}{B_v^2}\right)\cdot(\nabla\chi\times\nabla p) + O(\epsilon^3) = 0.\label{eq:divj}
\end{equation}
The only other equation of Strauss' reduced MHD that is nontrivial in the equilibrium limit can be written simply as $\BB\cdot\nabla p = 0$.

Equation \eqref{eq:divj} can be further simplified by exploiting the two-term quasisymmetry condition, the imposition of which is equivalent to demanding that $|\BB| = B(\psi, M\theta - N\phi)$ \citep{helander2014theory}: $(\BB\times\nabla\psi)\cdot\nabla B = F(\psi)\BB\cdot\nabla B$, where $\psi$ is the toroidal flux. Applying the ordering, this condition can be written in one of two equivalent ways:
\begin{subequations}
    \begin{gather}
        (\nabla\chi\times\nabla\Psi)\cdot\nabla B_v = \BB\cdot\nabla B_v,\label{eq:QSA}\\
		(\nabla B_v\times\nabla\chi)\cdot\nabla(\Psi + A) = \nabla\chi\cdot\nabla B_v,\label{eq:QSB}
    \end{gather}
\end{subequations}
where an alternate flux surface label $\Psi = \int d\psi/F(\psi)$ is defined to absorb the $F(\psi)$ factor. Note that $F(\psi)$ is $O(\epsilon^{-1})$ \citep{helander2014theory}, so $\Psi = O(\epsilon)$. Using \eqref{eq:QSA}, the second term in equation \eqref{eq:divj} can be rewritten as $p'(\nabla\chi\times\nabla\Psi)\cdot\nabla B_v^{-2} = p'\BB\cdot\nabla B_v^{-2}$, which results in both terms in the equation having $\BB\cdot\nabla$ acting on something. Since $p'$ commutes with $\BB\cdot\nabla$, we can remove the $\BB\cdot\nabla$ operator, obtaining a Grad-Shafranov-like equation:
\begin{equation}
    \Delta^* A - \mu_0\frac{dp}{d\Psi}\left(\frac{1}{B_v^2} - \frac{1}{B_0^2}\right) = H(\Psi),\label{eq:preGS}
\end{equation}
where $H$ and $\mu_0 p'/B_0^2$ are functions of $\Psi$ that appear due to the integration. Although $H$ is arbitrary, the $B_0$ term cannot be absorbed into it as it is necessary to cancel the $O(1)$ part of $1/B_v^2$ so that all terms are of the same order. Further, note that $H$ is a degree of freedom that allows us to specify the toroidal current profile; its axisymmetric equivalent is the $FF'$ term in the standard Grad-Shafranov equation.

To make further progress, $\chi$ has to be specified explicitly. The constraint $B_v = B_0 + O(\epsilon)$, which, as we have seen, arises from the high-$\beta$ assumption, requires that $\chi = B_0 l$. This can be seen by writing $B_v^2$ in orthogonal Mercier coordinates $(\rho,\omega,l)$ aligned to a magnetic axis with curvature $\kappa$ and torsion $\tau$ \citep{jorge2020construction,solovev1970reviews}:
\begin{equation*}
    B_v^2 = \left(\frac{\partial\chi}{\partial\rho}\right)^2 + \frac{1}{\rho^2}\left(\frac{\partial\chi}{\partial\omega}\right)^2 + \frac{1}{h^2}\left(\frac{\partial\chi}{\partial l}\right)^2 = B_0^2 + O(\epsilon),
\end{equation*}
where $h = 1 - \kappa\rho\cos\theta$, $\theta = \omega - \int_0^l\tau dl$. On the axis, we must have $\partial\chi/\partial\rho|_{\rho=0} = \lim_{\rho\to 0}\rho^{-1}\partial\chi/\partial\omega = 0$, since the axis is a field line. Thus, the constant $B_0^2$ can only come from the last term. It then follows that $\kappa\rho = O(\epsilon)$, so that $1/h^2 = 1 + O(\epsilon)$; otherwise $\chi$ would have depended on $\rho$ and $\omega$ at the lowest order, a contradiction. With that in mind, we can integrate at the zeroth order, obtaining $\chi = B_0 l + O(\epsilon)$. We can ignore the $O(\epsilon)$ correction to $\chi$ without loss of generality; as will be shown at the end of the section, it can be absorbed into the $\nabla A\times\nabla\chi$ term. Note that we can now give $\epsilon$ a more intuitive meaning: given that $\rho \sim r_0$ and $\kappa,\tau \sim 1/R_0$, we should order $r_0 = O(1)$ and $R_0 = O(\epsilon)$, hence we have a large aspect ratio expansion with $\epsilon$ acting as the inverse aspect ratio. Finally, it should be emphasized that the only arbitrary assumption made in this derivation is that $p = O(\epsilon)$. The rest of the ordering, while not rigorous, are still justified by heuristic arguments.

As a sanity check, we will demonstrate that $\chi = B_0 l$ only breaks the divergence-free condition at second order, which is acceptable, since this is a first-order model. Using the Laplace operator in orthogonal Mercier coordinates (equation (3.4) of \citet{sengupta2024stellarator}), we have:
\begin{equation*}
    \nabla^2\chi = \frac{1}{\rho h}\frac{\partial}{\partial\rho}\left(\rho h\frac{\partial\chi}{\partial\rho}\right) + \frac{1}{\rho^2 h}\frac{\partial}{\partial\omega}\left(h\frac{\partial\chi}{\partial\omega}\right) + \frac{1}{h}\frac{\partial}{\partial l}\left(\frac{1}{h}\frac{\partial\chi}{\partial l}\right) = -\frac{B_0}{h^3}\frac{\partial h}{\partial l} = O(\epsilon^2),
\end{equation*}
since $h = 1 + O(\epsilon)$ and $\partial/\partial l = B_0^{-1}h^2\nabla\chi\cdot\nabla = O(\epsilon)$.

Throughout the rest of the paper, we will use non-orthogonal Mercier coordinates $(x,y,l)$, where $x = \rho\cos\theta$ and $y = \rho\sin\theta$. The metric tensor is as follows \citep{solovev1970reviews}:
\begin{equation*}
    g^{ik} = \frac{1}{h^2}\begin{pmatrix}
        h^2 + \tau^2 y^2 & -\tau^2 xy & \tau y \\
        -\tau^2 xy & h^2 + \tau^2 x^2 & -\tau x \\
        \tau y & -\tau x & 1
    \end{pmatrix}, \qquad h = 1 - \kappa x.
\end{equation*}
Using the metric tensor and $\chi = B_0 l$, which implies $B_v = B_0/h$, equation \eqref{eq:QSB} can be written as
\begin{equation*}
    B_0^2\frac{\partial}{\partial y}(\Psi + A) = -B_0^2\left(\tau y + \frac{1}{\kappa}\frac{d\kappa}{dl}x\right).
\end{equation*}
The non-orthogonal Mercier coordinates have simplified the equation such that it can now be integrated:
\begin{equation}
    \Psi + A = -\frac{\tau y^2}{2} - \frac{1}{\kappa}\frac{d\kappa}{dl}xy + a(x,l),\label{eq:QSC}
\end{equation}
where $a$ is an arbitrary function. This allows us to write equation \eqref{eq:preGS} as an equation for $\Psi$:
\begin{equation}
    \Delta_\perp\Psi + \tau - \frac{\partial^2 a}{\partial x^2} = \frac{2\mu_0}{B_0^2}\frac{dp}{d\Psi}\kappa x - H(\Psi),\label{eq:GS}
\end{equation}
where we have used $B_v = B_0|\nabla l| = B_0/(1 - \kappa x)$ and $\Delta^* = \Delta_\perp + O(\epsilon)$ and dropped terms of order $\epsilon^2$ and higher.

Although we now have an equation for $\Psi$, solving it is nontrivial, as $\Psi$ is overconstrained, due to also having to satisfy the equation $\BB\cdot\nabla\Psi = 0$, which, after inserting $A$ from equation \eqref{eq:QSC}, can be written in Mercier coordinates as
\begin{equation}
    \frac{\partial\Psi}{\partial l}  - \frac{1}{\kappa}\frac{d\kappa}{dl}x\frac{\partial\Psi}{\partial x} + \left(\frac{1}{\kappa}\frac{d\kappa}{dl}y- \tau x - \frac{\partial a}{\partial x}\right)\frac{\partial\Psi}{\partial y}= 0,\label{eq:B.gpsM}
\end{equation}
where we have inserted the expression for $A$ from equation \eqref{eq:QSC}. The characteristics of this equation are given by
\begin{equation}
    \kappa x = C_1; \qquad \frac{y}{\kappa} = - C_1\int\frac{\tau}{\kappa^2}dl - \int\left.\frac{1}{\kappa}\frac{\partial a}{\partial x}\right|_{x=C_1/\kappa}dl + C_2.\label{eq:char}
\end{equation}
In order for $\Psi$ to satisfy equation \eqref{eq:B.gpsM}, it must be a function of only $C_1$ and $C_2$ as defined in \eqref{eq:char}. One must then ensure that the function also satisfies equation \eqref{eq:GS} at all values of $l$.

Having derived the model, we now show that adding an $O(\epsilon)$ correction to $\chi$ will not change it. Thus, if the assumption that the $B_0 l$ is the dominant term in $\chi$ holds, the model is fully general. Indeed, suppose that we add a correction $\chi_1 = O(\epsilon)$ to $\chi$, then at order $\epsilon$ this term can be absorbed into the $\nabla A\times\nabla\chi$ term by replacing $A\mapsto A + \widetilde{A}$, where $\nabla\chi_1 = B_0\nabla\widetilde{A}\times\nabla l$. The fact that such an $\widetilde{A}$ exists can be seen by writing out the contravariant $x$ and $y$ components of this relation:
\begin{equation*}
    \frac{\partial\chi_1}{\partial x} = B_0\frac{\partial\widetilde{A}}{\partial y} + O(\epsilon^2), \qquad\qquad \frac{\partial\chi_1}{\partial y} = -B_0\frac{\partial\widetilde{A}}{\partial x} + O(\epsilon^2).
\end{equation*}
These are just the Cauchy-Riemann equations; since $\chi_1$ must satisfy the Laplace equation at order $\epsilon$, a corresponding $\widetilde{A}$ is guaranteed to exist.

\section{Consistency with the first-order near-axis model}\label{sec:consistency}
The simplest ansatz for $\Psi$ that will work in this model is a quadratic polynomial in $C_1$ and $C_2$, which are defined in equation \eqref{eq:char}:
\begin{equation}
    \begin{aligned}
        \Psi &= s_1 C_1^2 + s_2 C_1 C_2 + s_3 C_2^2 = s_1 (\kappa x)^2 + s_2 (\kappa x)\left(\frac{y}{\kappa} + \kappa x\Lambda\right) + s_3 \left(\frac{y}{\kappa} + \kappa x\Lambda\right)^2 \\
        &= (s_1 + s_2\Lambda + s_3\Lambda^2)(\kappa x)^2 + (s_2 + 2s_3\Lambda)xy + s_3\left(\frac{y}{\kappa}\right)^2,
    \end{aligned}\label{eq:pspoly}
\end{equation}
where $\Lambda = \int\kappa^{-2}(\tau + a_2)dl$; and $a$ was chosen as $a = a_2(l)x^2/2$. This represents a plasma with an elliptic cross-section that rotates about the magnetic axis as $l$ varies. Inserting this expression into the Grad-Shafranov equation \eqref{eq:GS}, we obtain an equation expressing $a_2$ in terms of $\Lambda$. Note that, when inserting \eqref{eq:pspoly} into the Grad-Shafranov equation, all terms on the left-hand-side (LHS) of equation \eqref{eq:GS} will depend only on $l$, whereas terms on the RHS can depend on all three variables. Thus, we further assume $p = O(\epsilon^2)$ and $H(\Psi) = H_0 = \const$; this is consistent with the near-axis model, where pressure enters only at the second order. Proceeding, we have:
\begin{equation}
    2(s_1 + s_2\Lambda + s_3\Lambda^2)\kappa^2 + 2\frac{s_3}{\kappa^2} + \tau - a_2 + H_0 = 0.\label{eq:a2eq}
\end{equation}
Finally, we can combine $d\Lambda/dl = (\tau + a_2)/\kappa^2$, which follows from the definition of $\Lambda$, with the above equation, resulting in a Riccati equation for $\Lambda$:
\begin{equation}
    \frac{d\Lambda}{dl} - 2(s_1 + s_2\Lambda + s_3\Lambda^2) - \frac{2s_3}{\kappa^4} - \frac{2\tau}{\kappa^2} - \frac{H_0}{\kappa^2} = 0.\label{eq:riccati_L}
\end{equation}
The last step is to enforce periodicity on $\Lambda$. Averaging the above equation over $l$ removes the derivative term, resulting in the following constraint:
\begin{equation}
    -2(s_1 + s_2\langle\Lambda\rangle + s_3\langle\Lambda^2\rangle) - 2s_3\left\langle\frac{1}{\kappa^4}\right\rangle - 2\left\langle\frac{\tau}{\kappa^2}\right\rangle - H_0\left\langle\frac{1}{\kappa^2}\right\rangle = 0,\label{eq:s_i_cnstrt}
\end{equation}
where $\langle f(l)\rangle = L^{-1}\int_0^L f(l)dl$, with $L$ being the axis length. Thus, only two of the $s_i$ constants are free, with the remaining one determined by the above constraint. In practice, since $\Lambda$ is not known \emph{a priori}, equations \eqref{eq:riccati_L} and \eqref{eq:s_i_cnstrt} must be solved iteratively: first one makes an initial guess for the unknown $s_i$, then equation \eqref{eq:riccati_L} is solved for $\Lambda$, which allows one to calculate the corrected value of the unknown $s_i$ from equation \eqref{eq:s_i_cnstrt}; this cycle is repeated until convergence is achieved. This same method is used in the near-axis model to solve the $\sigma$-equation while simultaneously finding the correct rotational transform that is compatible with periodicity \citep{landreman2019direct}.

Now consider the lowest order near-axis expression for $\Psi$ \citep{jorge2020construction}:
\begin{equation}
    \begin{aligned}
        \Psi &\simeq \frac{\psi}{F_0} \simeq \frac{\pi B_0}{F_0}\rho^2[\cosh\eta + \sinh\eta\cos 2(\theta + \delta)] \\
        &= \frac{\pi B_0}{F_0}[(\cosh\eta + \sinh\eta\cos 2\delta)x^2 - 2\sinh\eta\sin 2\delta~xy +(\cosh\eta - \sinh\eta\cos 2\delta)y^2],
    \end{aligned}\label{eq:psnae}
\end{equation}
where $x = \rho\cos\theta$, $y = \rho\sin\theta$, $F_0$ is the value of $F(\psi)$ on axis and $\eta$ and $\delta$ are functions of $l$. Note that $F_0$ as defined here is the inverse of the $F_0$ in \citet{jorge2020construction}. In quasisymmetry, we have $\cosh\eta - \sinh\eta\cos 2\delta = \etabar^2/\kappa^2$ \citep{jorge2020construction}. Since $\etabar = \const$\footnote{In the near-axis literature, $\etabar$ is a parameter that controls the extent of the flux surfaces in the normal direction for a given curvature. The maximum $x$ for a flux surface $\psi$ is $x_{max}(l) = \etabar\sqrt{\psi}/(\kappa(l)\sqrt{\pi B_0})$; see Figure 2 in \citet{rodriguez2021solving} for an illustration.}, the $y^2$ term of \eqref{eq:pspoly} agrees with that of \eqref{eq:psnae} and $s_3 = \pi B_0\etabar^2/F_0$. Further, taking $4(\text{coef. of }x^2)(\text{coef. of }y^2) - (\text{coef. of } xy)^2$ from \eqref{eq:psnae}, we see that the $l$-dependent terms cancel and we are left with just a constant, $4\pi^2 B_0^2/F_0^2$. Likewise, when using the corresponding coefficients from \eqref{eq:pspoly}, the terms with $\Lambda$ cancel and we get $4s_1 s_3 - s_2^2 = 4\pi^2 B_0^2/F_0^2$.

Next, we show that equation \eqref{eq:riccati_L} is equivalent to the $\sigma$-equation in \citet{jorge2020construction}. A relation between $\Lambda$ and $\sigma$ can be obtained by comparing the coefficients of $xy$ in equations \eqref{eq:pspoly} and \eqref{eq:psnae}. Since $\sigma = \sinh\eta\sin 2\delta$ \citep{jorge2020construction}, we have
\begin{equation*}
    \Lambda = -\frac{\pi B_0}{F_0 s_3}\sigma - \frac{s_2}{2s_3}.
\end{equation*}
Inserting this into equation \eqref{eq:riccati_L} and replacing the $s_i$'s with near-axis variables, we get
\begin{equation}
    \frac{d\sigma}{dl} + \frac{2\pi B_0}{F_0} \left(1 + \sigma^2 + \frac{\etabar^4}{\kappa^4}\right) + 2\tau\frac{\etabar^2}{\kappa^2} + H_0\frac{\etabar^2}{\kappa^2} = 0.\label{eq:riccati_s}
\end{equation}
In the case when $H_0 = 0$, this equation reduces to the equation (29) in \citet{jorge2020construction}.

To conclude this section, we show that when $s_2 = 0$ there is a simple relationship between the characteristics and Boozer angles. Choosing a value for $s_2$ is similar to gauge fixing, since a nonzero $s_2$ simply represents a deviation of the ellipse from the upright position in the $(C_1,C_2)$ plane. The orientation of the ellipse in real space is determined by $\sigma$, which is unaffected by changes in $s_2$, as long as the inital condition for $\Lambda$ is changed accordingly. The coordinates $x$ and $y$ can be written in terms of $C_1,C_2$ as
\begin{equation}
    x = \frac{C_1}{\kappa}, \qquad y = \kappa C_2 - \kappa\Lambda C_1.
\end{equation}
Meanwhile, equations (1) and (54) from \citet{jorge2020construction} give
\begin{equation}
    \boldsymbol{r} = \boldsymbol{r}_0 + x\boldsymbol{n} + y\boldsymbol{b} = \boldsymbol{r}_0 + \sqrt{\frac{\psi}{\pi B_0}}\left[\frac{\etabar}{\kappa}\cos\vartheta\boldsymbol{n} + \frac{\kappa}{\etabar}(\sin\vartheta + \sigma\cos\vartheta)\boldsymbol{b}\right].
\end{equation}
Equating the two and using the relations between $\Lambda$ and $\sigma$ as well as $s_3$ and $\etabar$, we arrive at the following:
\begin{equation}
    C_1 = \frac{F_0}{\pi B_0}\sqrt{s_3\Psi}\cos\vartheta, \qquad C_2 = \sqrt{\frac{\Psi}{s_3}}\sin\vartheta,
\end{equation}
where $\vartheta = \theta_B - N\phi_B$, with $\theta_B$ and $\phi_B$ being the Boozer angles. 

\section{New classes of solutions}\label{sec:newsoln}
We now present three classes of exact and approximate solutions to equations \eqref{eq:GS} and \eqref{eq:B.gpsM} that fall outside the scope of the near-axis model. We will begin with a cubic polynomial solution with constant $H$ and then proceed to a non-polynomial solution and a solution with variable $H$.

\subsection{Cubic solution with finite pressure}\label{sec:newsoln:highbeta}
Consider the following cubic polynomial in $C_1$ and $C_2$:
\begin{equation}
    \begin{aligned}
        \Psi &= s_1 C_1^2 + s_2 C_1 C_2 + s_3 C_2^2 + r C_1^3 \\
        &= (s_1 + s_2\Lambda + s_3\Lambda^2)(\kappa x)^2 + (s_2 + 2s_3\Lambda)xy + s_3\left(\frac{y}{\kappa}\right)^2 + r(\kappa x)^3.\label{eq:pspoly3}
    \end{aligned}
\end{equation}
Inserting this ansatz into equation \eqref{eq:GS}, the following is obtained:
\begin{equation}
    2(s_1 + s_2\Lambda + s_3\Lambda^2)\kappa^2 + \frac{2s_3}{\kappa^2} + 6r\kappa^3 x + \tau - a_2 = \frac{2\mu_0 p_1}{B_0^2}\kappa x - H_0,\label{eq:a2req}
\end{equation}
where we have additionally assumed that $dp/d\Psi = p_1 = \const$, $H(\Psi) = H_0 = \const$ and $a = a_2(l)x^2/2$. Only two terms in the above equation depend on $x$; equating those two terms, we get
\begin{equation}
    r = \frac{\mu_0 p_1}{3B_0^2\kappa^2},\label{eq:req}
\end{equation}
while the remaining terms match equation \eqref{eq:a2eq}. Thus, a solution is obtained by first finding a quadratic solution as discussed in Section \ref{sec:consistency}, and then adding a cubic term $r(\kappa x)^3$ with $r$ given by equation \eqref{eq:req}. Equation \eqref{eq:req} also imposes the constraint that $\kappa = \const$, but, since $\tau$ is unconstrained, we can still get nonplanar closed curves. At this point, we can see why $rC_1^3$ is the only cubic term that we can include in \eqref{eq:pspoly3}. If we were to add cubic terms that involves $C_2$, then equation \eqref{eq:a2req} would also have terms that depend on $y$ and we would end up with an additional equation that would constrain $\tau$, meaning that we would not be able to ensure that the axis is closed.

Finally, note that, while this cubic solution seems to be superficially similar to a second order near-axis solution, there is a fundamental difference. Namely, the near-axis model assumes that the cubic term is a correction that is much smaller than the quadratic terms, whereas the present solution allows the cubic term to be of the same order as the quadratic ones. Also, unlike the second-order near-axis model, where the quasisymmerty error is $O(\epsilon^3)$, the present solution is still first order, so the quasisymmetry error will be $O(\epsilon^2)$.

\subsection{Non-polynomial approximate solutions}\label{sec:newsoln:nonpoly}
In this subsection, we will show an approximate solution that consists of a rotating ellipse, which, as we have seen in the previous section, can be represented as a quadratic polynomial, and a non-polynomial perturbation. As we will perform a subsidiary expansion, it is convenient to rescale all quantities to be zeroth order in $\epsilon$. Thus, after performing the following replacement: $\{\kappa,\etabar,\tau,a_2,H_0,\Psi,d/dl\}\mapsto\epsilon\{\kappa,\etabar,\tau,a_2,H_0,\Psi,d/dl\}$ and $F_0\mapsto F_0/\epsilon$, we see that the $\epsilon$ parameter is canceled in equation \eqref{eq:GS} and the rotating ellipse solution \eqref{eq:pspoly}.

Now consider the case when $s_2 = 0$; then, using the expressions for the $s_i$'s and $\Lambda$ that we found in section \ref{sec:consistency}, we have $s_3 = \pi B_0\etabar^2/F_0$, $s_1 = \pi B_0/(F_0\etabar^2)$ and $\Lambda = -\sigma/\etabar^2$. The rotating ellipse solution can be represented as
\begin{equation}
    \Psi_{re} = \frac{\pi B_0}{F_0}\left(\frac{1}{\etabar^2}C_1^2 + \etabar^2 C_2^2\right) = \frac{\pi B_0}{F_0}(c_1^2 + c_2^2),\label{eq:re_nf}
\end{equation}
where the integration constants of the characteristics \eqref{eq:char} have been rescaled with respect to $\etabar$:
\begin{equation}
    c_1 = \frac{C_1}{\etabar} = \frac{\kappa x}{\etabar},\qquad\qquad c_2 = \etabar C_2 = \frac{\etabar y}{\kappa} + \etabar\kappa x\Lambda = \frac{\etabar y}{\kappa} - \frac{\kappa x\sigma}{\etabar}
\end{equation}
yielding the normal form representation. We add to \eqref{eq:re_nf} a non-polynomial perturbation $f$, and use the relation $F_0^{-1} = (\iota_0 - N)/(2\pi B_0 r_0)$, which was found in \citet{jorge2020construction}, with $\iota_0$ being the rotational transform on axis and $N$ the helicity of the quasisymmetry. Here, the axis length $L$ was replaced with the characteristic value of the minor radius $r_0 = \epsilon L/(2\pi)$, due to $F_0$ having been rescaled with respect to $\epsilon$. The resulting ansatz is as follows:
\begin{equation}
    \Psi = \frac{\iota_0 - N}{2r_0}\left[(c_1^2 + c_2^2) + \left(\frac{\etabar}{\kappa_0}\right)^2 f(c_2 + \sigma_0 c_1)\right],\label{eq:pstw}
\end{equation}
where $\kappa_0$ is the minimum value of $\kappa$. Inserting this ansatz into the Grad-Shafranov equation \eqref{eq:GS}, multiplying everything by $(\etabar/\kappa)^2$ and using $a_2 = \kappa^2 d\Lambda/dl - \tau = -(\kappa/\etabar)^2 d\sigma/dl - \tau$, the following is obtained:
\begin{equation}
    \begin{aligned}
        \frac{\iota_0 - N}{2r_0}&\left[2(1 + \sigma^2) + 2\left(\frac{\etabar}{\kappa}\right)^4 + \left(\frac{\etabar}{\kappa_0}\right)^2\left((\sigma_0 - \sigma)^2 + \left(\frac{\etabar}{\kappa}\right)^4\right)f''\right] \\
        &+ \frac{d\sigma}{dl} + \left(\frac{\etabar}{\kappa}\right)^2(2\tau + H_0) = 0.\label{eq:pert_riccati_s}
    \end{aligned}
\end{equation}
We can now carry out a subsidiary expansion in the limit of small $\etabar^2$, i.e. $(\etabar/\kappa_0)^2 \ll 1$, but $(\etabar/\kappa_0)^4 > \epsilon$ since we keep terms at that order in \eqref{eq:pstw}. This corresponds to the limit of a high aspect ratio stellarator with a highly elongated cross section. Given that $\sigma\sim\etabar^2$ and $\iota_0 - N \sim\etabar^2$ \citep{rodriguez2022quasisymmetry}, the two lowest orders of the above equation are as follows:
\begin{equation}
    \begin{aligned}
        &O\left(\left(\frac{\etabar}{\kappa_0}\right)^2\right):    &&\frac{d\sigma_2}{dl} + \frac{\iota_{0,2}}{r_0} = -\left(\frac{\etabar}{\kappa}\right)^2(2\tau + H_0), \\
        &O\left(\left(\frac{\etabar}{\kappa_0}\right)^6\right):    &&\frac{d\sigma_6}{dl} + \frac{\iota_{0,6}}{r_0} = -\frac{\iota_{0,2}}{r_0}\left[\sigma_2^2 + \left(\frac{\etabar}{\kappa}\right)^4\right],
    \end{aligned}\label{eq:sigeq26}
\end{equation}
where $\sigma = \sigma_2 + \sigma_6$ with $\sigma_n\sim\etabar^n$, and $\iota_0 = N + \iota_{0,2} + \iota_{0,6}$ with $\iota_{0,n}\sim\etabar^n$. The values of $\iota_{0,n}$ are determined by enforcing periodicity on $\sigma_n$: equations \eqref{eq:sigeq26} are averaged over $l$, which removes $d\sigma_n/dl$; the $\iota_{0,n}$ term must then be equal to the average of the RHS. Since we only solve equation \eqref{eq:pert_riccati_s} up to $O((\etabar/\kappa_0)^6)$ and $f$ does not appear until $O((\etabar/\kappa_0)^8)$, we can treat $f$ as a free function. If we were to attempt to solve this equation at $O((\etabar/\kappa_0)^8)$, then only the trivial solution $f'' = \const$ (i.e. $f$ is a quadratic polynomial) would be permitted.

\subsection{Approximate solution with variable $H$}\label{sec:varH}
We wrap up this section by presenting a rotating ellipse approximate solution where $H(\Psi) = H_0 + H_1\Psi$. We also add a quartic term to $a$: $a = a_2(l)x^2/2 + a_4(l)x^4/24$, so the rescaled characteristics become
\begin{equation}
    \begin{gathered}
        c_1 = \frac{\kappa x}{\etabar}, \qquad c_2 = \frac{\etabar y}{\kappa} + \etabar\kappa x\Lambda + \frac{\etabar\kappa^3 x^3}{6}V, \\
        \Lambda = \int\frac{\tau + a_2}{\kappa^2}dl, \qquad V = \int\frac{a_4}{\kappa^4}dl.
    \end{gathered}\label{eq:a4char}
\end{equation}
Just like in the previous subsection, we rescale all quantities to be zeroth order in $\epsilon$ and let $s_2 = 0$. In addition, we order $H_1\sim\etabar^2$ and $a_4\sim\etabar^2$. It can now be seen that in the rotating ellipse ansatz \eqref{eq:re_nf} terms with $x$ or $y$ raised to a power higher than two only appear at order $(\etabar/\kappa_0)^6$ and higher. Thus, up to $O((\etabar/\kappa_0)^4)$, it is still purely a rotating ellipse.

Inserting the rotating ellipse solution \eqref{eq:re_nf} with the new characteristics \eqref{eq:a4char} into the Grad-Shafranov equation \eqref{eq:GS} and multiplying everything by $(\etabar/\kappa)^2$, the following equation is obtained:
\begin{equation}
    \begin{aligned}
        &\frac{\iota_0 - N}{r_0}\left[(1 + \sigma^2) + \left(\frac{\etabar}{\kappa}\right)^4\right] + \frac{d\sigma}{dl} + \left(\frac{\etabar}{\kappa}\right)^2\left(2\tau - a_4\frac{x^2}{2} + H_0\right) \\
        &+ \frac{\iota_0 - N}{2r_0}H_1 x^2 + O\left(\left(\frac{\etabar}{\kappa_0}\right)^8\right) = 0,
    \end{aligned}
\end{equation}
where we have again used $a_2 = -(\kappa/\etabar)^2 d\sigma/dl - \tau$. Just as in the previous subsection, the above equation can be solved order by order. The three lowest orders are as follows:
\begin{equation}
    \begin{aligned}
        &O\left(\left(\frac{\etabar}{\kappa_0}\right)^2\right):    &&\frac{d\sigma_2}{dl} + \frac{\iota_{0,2}}{r_0} = -\left(\frac{\etabar}{\kappa}\right)^2(2\tau + H_0), \\
        &O\left(\left(\frac{\etabar}{\kappa_0}\right)^4\right):    &&\left(\frac{\etabar}{\kappa}\right)^2 a_4 = \frac{\iota_{0,2}}{r_0}H_1, \\
        &O\left(\left(\frac{\etabar}{\kappa_0}\right)^6\right):    &&\frac{d\sigma_6}{dl} + \frac{\iota_{0,6}}{r_0} = -\frac{\iota_{0,2}}{r_0}\left[\sigma_2^2 + \left(\frac{\etabar}{\kappa}\right)^4\right].
    \end{aligned}\label{eq:sigeq246}
\end{equation}
Thus, $\sigma$ is still determined by the same equations as \eqref{eq:sigeq26}; in addition to that, we also have an equation to determine $a_4$.

\section{Numerical example}\label{sec:numref}
While the model presented in this paper was derived with high-$\beta$ equilibria in mind, the only high-$\beta$ equilibrium that we have been able to find analytically is the constant curvature one presented in section \ref{sec:newsoln:highbeta}. In this section, we will instead illustrate the approximate solution discussed in section \ref{sec:newsoln:nonpoly} with a numerical example of a quasiaxisymmetric ($N=0$) device. This approximate solution has another important property that we wanted our model to include: nonzero shear. We were unable to get a numerical equilibrium with high $\beta$ because VMEC failed to find a solution for all of the nonplanar constant curvature axes that we could find, due to their high shaping. We leave the numerical verification of the analytical solution in section \ref{sec:newsoln:highbeta}, as well as the numerical search for other high-$\beta$ solutions for future work.

The example is constructed by numerically solving equations \eqref{eq:sigeq26}, and is then compared to both VMEC \citep{hirshman1983steepest} results and the closest near-axis approximation, which is identical to taking $f=0$. The pyQSC code \citep{pyQSC} was used to solve the $\sigma$ equations and generate VMEC input files, based on the method described in \citet{landreman2019constructing}.

\begin{figure}
    \centering
    \includegraphics[scale=0.35]{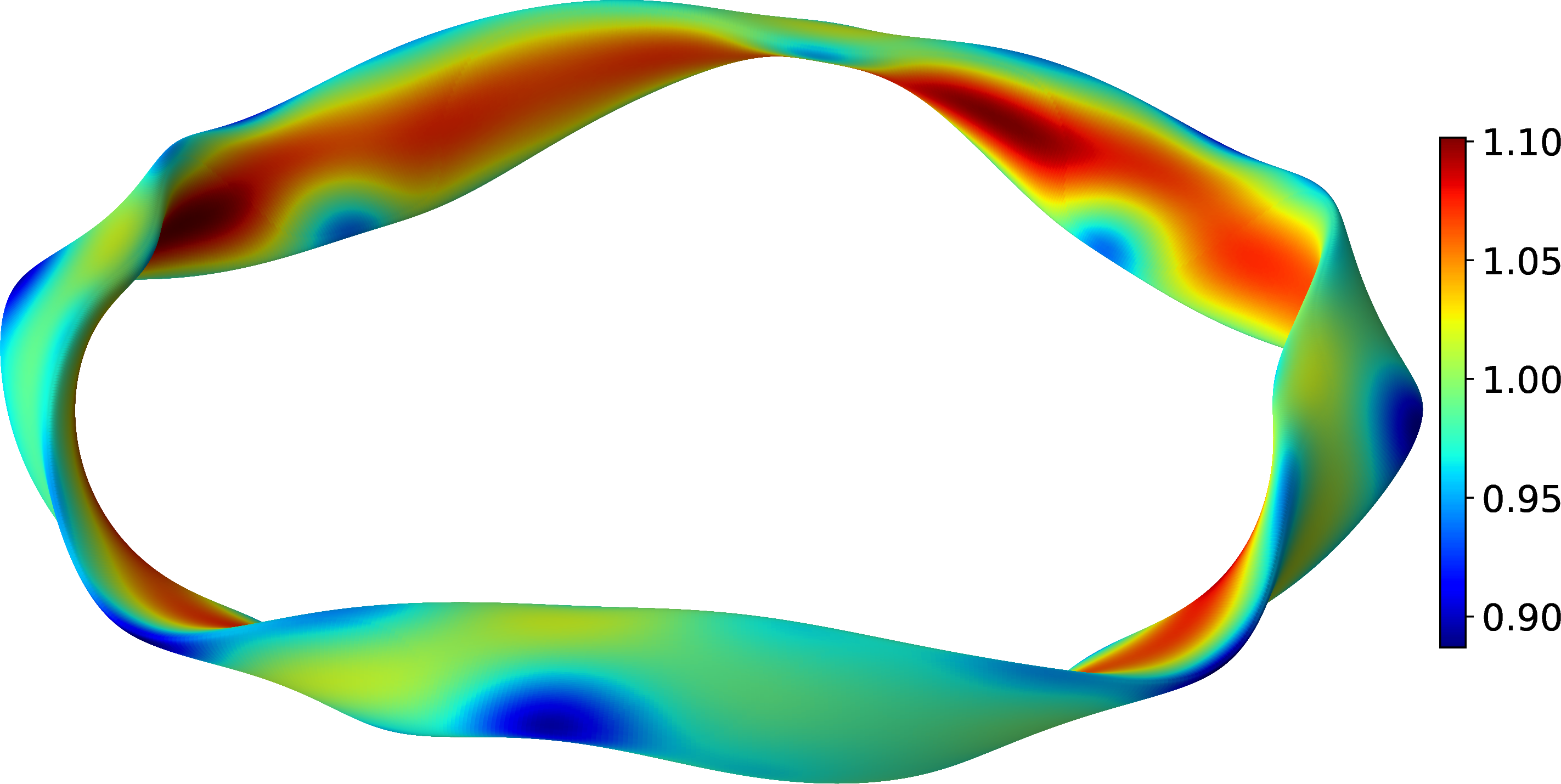}
    \caption{The outermost flux surface of the VMEC equilibrium based on the solution with a sixth order polynomial perturbation. Color represents $|\BB|$.}
    \label{fig:6pert_3D}
\end{figure}

While the solution in section \ref{sec:newsoln:nonpoly} allows for arbitrary $f$, here we will consider a sixth order polynomial, $f(c) = (0.3~\mathrm{m^{-2}})c^4 + (0.3~\mathrm{m^{-4}})c^6$, where $c = c_2 + \sigma_0 c_1$; this is higher order than the cubic polynomials of the second order near-axis model. We consider a four-field-period solution with $\etabar = 2.01735426\cdot 10^{-2}~\mathrm{m^{-1}}$, $B_0 = 1$~T, $H_0 = 0$. Equations \eqref{eq:sigeq26} result in rotational transforms $\iota_{0,2} = 0.188923174$ and $\iota_{0,6} = -0.031115615$. The axis shape given by:
\begin{equation}
    \begin{aligned}
        R(\phi) &= 29.7794783 - 3.63597602\cdot 10^{-1}\cos 4\phi + 1.47477208\cdot 10^{-1}\cos 8\phi \\
        &+ 1.35576435\cdot 10^{-2}\cos 12\phi, \\
        z(\phi) &= 1.93173817\sin 4\phi + 2.38762327\cdot 10^{-2}\sin 8\phi - 7.72243217\cdot 10^{-3}\sin 12\phi,
    \end{aligned}\label{eq:axis}
\end{equation}
where $R$ and $z$ are the cylindrical coordinates in meters. The value of $(\etabar/\kappa_0)^2$ for this axis is approximately 0.3. The VMEC equilibrium, obtained by constructing the $F_0\Psi/\pi = 4~\mathrm{T\cdot m^2}$ surface using equation \eqref{eq:pstw} and passing it to VMEC as the boundary, is shown in Figure \ref{fig:6pert_3D}. The aspect ratio of the resulting stellarator is 16.5. The assumption $dp/d\Psi = 0$, made when constructing the base solution in section \ref{sec:consistency}, and the choice of $H_0$ correspond to pressure and toroidal current profiles of zero; this was specified in the VMEC input. The total toroidal flux in the VMEC input was determined by multiplying $B_0$ by the plasma cross section area in the plane perpendicular to the axis.

\begin{figure}
    \centering
    \raisebox{-0.5\height}{\includegraphics[scale=0.65]{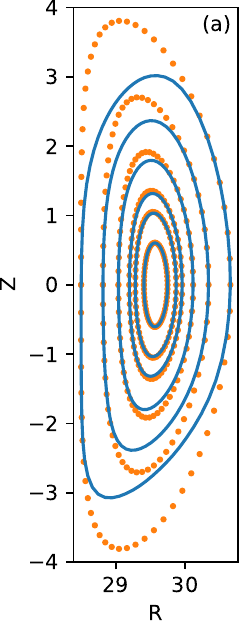}
    \includegraphics[scale=0.65]{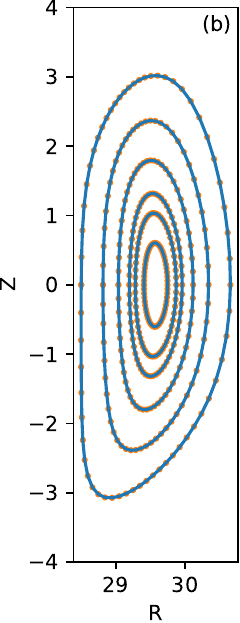}}\hfill
    \raisebox{-0.5\height}{\includegraphics[scale=0.49]{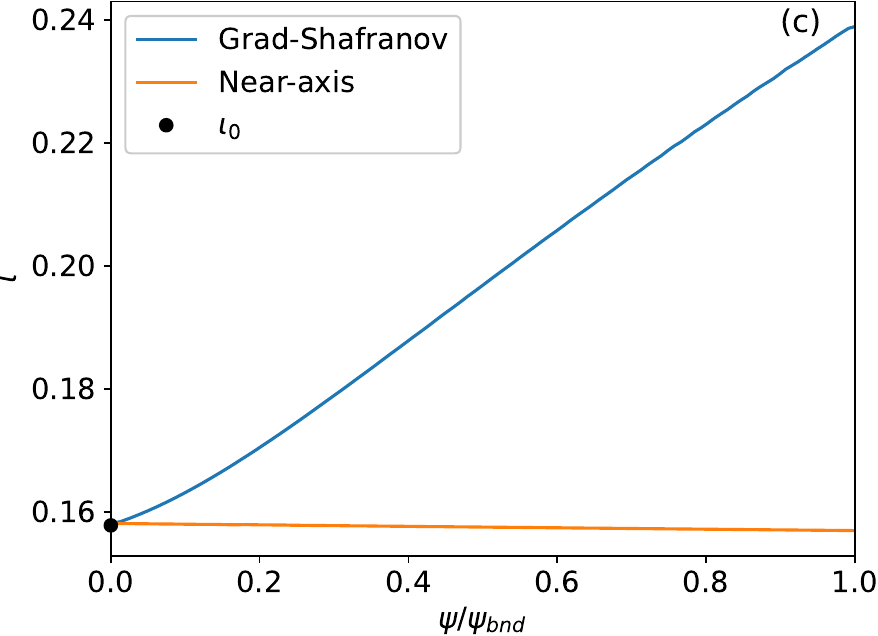}}
    \caption{A comparison of the flux surfaces computed from equation \eqref{eq:pstw} (lines) to the closest near-axis approximation (dots) is shown in (a). In (b), the flux surfaces computed from equation \eqref{eq:pstw} (lines) are compared to the flux surfaces computed by VMEC (dots). The flux surfaces are shown for $F_0\Psi/\pi = 0.1, 0.3, 0.5, 1, 2, 4~\mathrm{T\cdot m^2}$. Finally, (c) shows the $\iota$ profiles computed in the Grad-Shafranov model, the near-axis model and the constant $\iota_0$.}
    \label{fig:comparison}
\end{figure}

\begin{figure}
    \centering
    \includegraphics[scale=0.5]{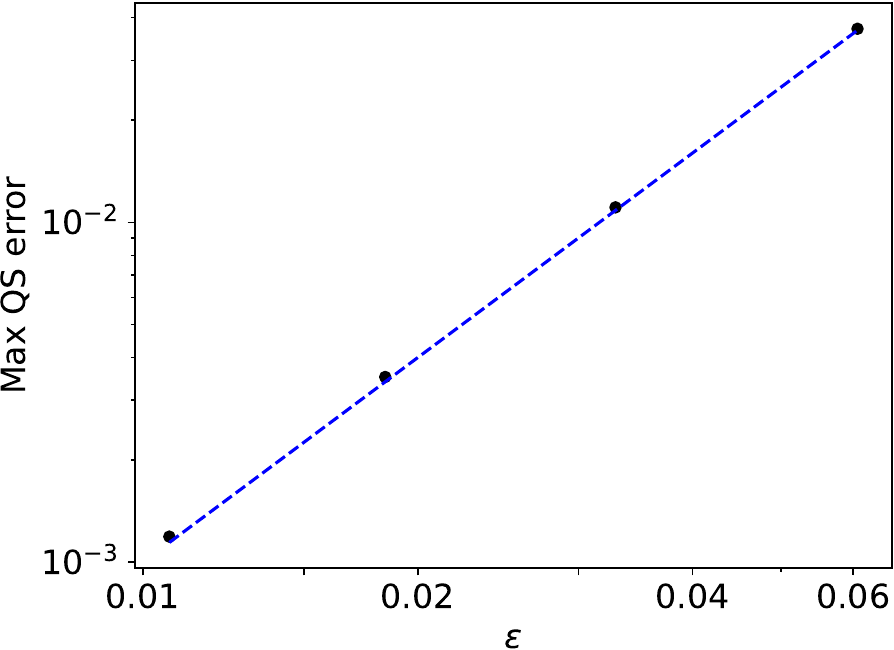}
    \caption{The maximum quasisymmetry error (black dots) scales as $\epsilon^2$ (dashed blue line is $10\epsilon^2$).}
    \label{fig:QSe}
\end{figure}

Figures \ref{fig:comparison}a,b compare the flux surfaces of the present Grad-Shafranov model, obtained from equation \eqref{eq:pstw}, to both the flux surfaces calculated by VMEC and the flux surfaces of the closest near-axis approximation, as given by equation \eqref{eq:re_nf}. All flux surfaces are shown on the $\phi = 0$ poloidal plane. As expected, near the axis, the near-axis flux surfaces closely match those of the Grad-Shafranov model, but further away from the axis the Grad-Shafranov surfaces become non-elliptic as the contribution from $f$ becomes non-negligible. Figure \ref{fig:comparison}c compares the $\iota$ profiles in the near-axis and Grad-Shafranov models. These are computed numerically by first calculating the toroidal flux $\psi$ at each value of $\Psi$, then finding $F = d\Psi/d\psi$ and using the formula
\begin{equation}
    F(\psi) = \frac{G(\psi) + NI(\psi)}{\iota(\psi) - N},
\end{equation}
which is given in \citet{helander2014theory} as an unnumbered equation. Again, these agree well with each other and the $\iota_0$ constant near the axis, but the Grad-Shafranov solution has a noticeable shear. The $\iota$ profiles calculated by VMEC (not shown here) do not match the Grad-Shafranov and near-axis profiles shown in Figure \ref{fig:comparison}c, with the $\iota$ on axis in the corresponding VMEC solutions being greater by about 0.06 and 0.024, respectively. A similar mismatch with VMEC has been observed in \citet{sengupta2024stellarator}; it likely appears because the boundary (and not the axis) is fixed in VMEC, so for a finite aspect ratio, the VMEC axis does not match the axis given by \eqref{eq:axis}. There are also additional complications arising from VMEC having a coordinate singularity on the axis \citep{panici2023the}. Just as in \citet{sengupta2024stellarator}, the agreement between the $\iota$ on axis in VMEC and that predicted by the Grad-Shafranov model improves as the aspect ratio is increased. Finally, Figure \ref{fig:QSe} shows that the maximum quasisymmetry error, defined as
\begin{equation*}
    \max_\psi \sqrt{\sum_{n\neq 0}\widehat{B}_{n,m}(\psi)^2\Big/\sum_{n,m}\widehat{B}_{n,m}(\psi)^2},
\end{equation*}
and calculated in VMEC equilibria with varying aspect ratios, scales as $\epsilon^2$. Here, $\widehat{B}_{n,m}(\psi)$ are the Fourier modes of $|\BB|$ on flux surface $\psi$. This behavior is to be expected, since we have only ensured quasisymmetry at order $\epsilon$.

\section{Conclusion}
We have derived a first-order asymptotic model for high-$\beta$ ($\beta = O(\epsilon)$) quasisymmetric stellarators under the assumption that the vacuum magnetic field is dominant ($\epsilon \ll 1$), but non-vacuum effects can still contribute ($\nabla\chi\cdot\nabla = O(\epsilon)$). We have shown that these assumptions require the aspect ratio to be large and a simple expression is obtained for the lowest-order vacuum field ($\nabla\chi = B_0\nabla l$). The first-order correction, which involves both vacuum and current-driven fields, is then governed by a Grad-Shafranov equation and the requirement that flux surfaces exist ($\BB\cdot\nabla\Psi = 0$). Unlike the more simple near-axisymmetric case considered in our previous work \citep{sengupta2024asymptotic}, which is a special case of the present model, the overconstraining problem cannot be resolved in general. Thus, one must look for special solutions that satisfy both equations simultaneously. One family of such solutions, and thus another special case of this model, are the first-order quasisymmetric near-axis solutions.

We also provide several new solutions which are outside the scope of both the near-axis model and our previous near-axisymmetric model, and show a numerical example of one of these new solutions. When comparing the numerical results from the present model to VMEC, the flux surfaces match well, but the $\iota$ profiles show a mismatch which decreases as the aspect ratio is increased. The mismatch is most likely due to VMEC shifting the axis during minimization, so the VMEC axis does not exactly match the axis used when solving the Grad-Shafranov equation. We plan to use DESC in future work to re-evaluate the agreement between $\iota$ profiles. DESC has the option of fixing the axis \citep{panici2023near} and avoids large errors on axis, which are typical for VMEC, by using Zernike polynomials \citep{panici2023the}. Finally, DESC can handle highly shaped axes better than VMEC, and is a more appropriate tool for doing a numerical verification of the analytical solution in section \ref{sec:newsoln:highbeta}, as all of the nonplanar constant curvature axes that we have found so far are highly shaped.

We expect that the analytical solutions found in this paper are only a small subset of all possible solutions to the model we derived. In future work, we plan to implement a numerical solver that will look for solutions by constructing a function basis in $C_1,C_2$ space and looking for functions and boundary shapes that minimize the residual in the Grad-Shafranov equation \eqref{eq:GS}.

Another line of work that we intend to pursue is the construction of a similar model in the low-$\beta$ ($\beta = O(\epsilon^2)$) regime. As discussed in section \ref{sec:deriv}, in the low-$\beta$ regime, there are no constraints on $\chi$, aside from satisfying the Laplace equation. This allows for a richer set of solutions, including compact stellarators, at the expense of analytical progress being more difficult. Nevertheless, as will be shown in a future publication, some analytical progress, such as finding the characteristics of the $\BB\cdot\nabla\Psi = 0$ equation, is still possible. Finally, the low-$\beta$ model is closely related to the reduced MHD model implemented in the JOREK code \citep{hoelzl2020the,nikulsin2021models,nikulsin2021testing}, with the main difference being that the JOREK models use an ansatz instead of an ordering. Implementing a numerical solver for this model will allow one to initialize JOREK simulations of quasisymmetric stellarators from the Grad-Shafranov solutions instead of importing GVEC equilibria. \\

The authors thank Per Helander, Stefan Buller, Elizabeth Paul, Eduardo Rodriguez, Andrew Brown, Stuart Hudson, Matt Landreman, Felix Parra Diaz, Hongxuan Zhu and Vinicius Duarte for fruitful discussions.

This research was supported by a grant from the Simons Foundation/SFARI (560651, AB), and the Department of Energy Award No. DE-SC0024548.

\appendix
\section{Perpendicular force balance terms with general $B_v$}\label{sec:perp_fb}
In this appendix, we will provide a rigorous proof that the two terms in equation \eqref{eq:perp_fb} are linearly independent when $B_v \neq B_0 + O(\epsilon)$. First, suppose that the two terms are not linearly independent; then we can write
\begin{equation}
    \nabla\left(\frac{B_1}{\mu_0 B_v}\right) + \frac{1}{B_v^2}\nabla p = f\nabla g,\label{eq:lindep}
\end{equation}
where $f$ and $g$ are some functions to be determined. If the above equation holds, then the LHS should be orthogonal to its own curl:
\begin{equation}
    0 = \left[\nabla\left(\frac{B_1}{\mu_0 B_v}\right) + \frac{1}{B_v^2}\nabla p\right]\cdot\nabla\times\left[\nabla\left(\frac{B_1}{\mu_0 B_v}\right) + \frac{1}{B_v^2}\nabla p\right] = \frac{2}{\mu_0 B_v^4}\nabla B_1\cdot(\nabla p\times\nabla B_v).
\end{equation}
Thus, $B_1$ must be a function of only $p$ and $B_v$. Using this fact and dotting equation \eqref{eq:lindep} with $\nabla p\times\nabla B_v$, we get $\nabla g\cdot(\nabla p\times\nabla B_v) = 0$ if $f \neq 0$; thus $g$ must also be a function of only $p$ and $B_v$.

In order for equation \eqref{eq:perp_fb} to be satisfied, we must have either $f=0$ or $g=\const$ or $g = g(\chi)$. Consider the latter case first. Since $g(\chi) = g(p,B_v)$, this is an equation that can be solved for $B_v$, giving $B_v = B_v(p,\chi)$. If we insert this into the quasisymmetry condition \eqref{eq:QSA}, we will get $\BB\cdot\nabla\chi~\partial B_v/\partial\chi = 0$ since $\BB\cdot\nabla p = 0$, meaning that $B_v$ must be a flux function, a contradiction since $|\BB| = B_v + O(\epsilon^2)$ and $|\BB|$ can only be a flux function in axisymmetry \citep{schief2003nested}.

Alternatively, if either $f=0$ or $g=\const$, equation \eqref{eq:lindep} will have the following two components in the $\nabla p$ and $\nabla B_v$ directions:
\begin{equation}
    \frac{1}{\mu_0 B_v}\frac{\partial B_1}{\partial p} + \frac{1}{B_v^2} = 0, \qquad \frac{1}{\mu_0 B_v}\frac{\partial B_1}{\partial B_v} - \frac{B_1}{\mu_0 B_v} = 0.
\end{equation}
These two equations for $B_1$ are incompatible. Integrating the first one, we get $B_1/\mu_0 = -p/B_v + u(B_v)$, where $u$ is an arbitrary function. Inserting this into the second equation, we get $u' - u/B_v = -2p/B_v^2$, a contradiction since $u$ cannot depend on $p$. Finally note that the contradiction is resolved if $B_v = B_0 + O(\epsilon)$, since $\nabla B_v = O(\epsilon)$ in that case, which will remove the second equation, while the solution to the first equation will become the pressure-balance relation \eqref{eq:theta_fb}.

\section{Near-axisymmetric and near-helically-symmetric solutions}
In \citet{sengupta2024asymptotic}, we derived a condition for consistency between the Grad-Shafranov equation and the $\BB\cdot\nabla\Psi = 0$ equation, and then used it to obtain a condition on $a$ under which the two equations are consistent for all solutions $\Psi$. We can attempt a similar approach for the present model, but it will exclude many solutions of interest, including all solutions discussed in sections \ref{sec:consistency} and \ref{sec:newsoln}.

To obtain the consistency condition, we apply the $\BB\cdot\nabla$ operator to equation \eqref{eq:GS}, and commute $\BB\cdot\nabla$ with $\Delta_\perp$:
\begin{equation}
    \Delta_\perp\BB\cdot\nabla\Psi + [\BB\cdot\nabla,\Delta_\perp]\Psi + \BB\cdot\nabla\left(\tau - \frac{\partial^2 a}{\partial x^2} - \frac{2\mu_0}{B_0^2}\frac{dp}{d\Psi}\kappa x\right) = 0.
\end{equation}
The whole equation must be satisfied if equation \eqref{eq:GS} is satisfied; however the first term must be individually zero since $\BB\cdot\nabla\Psi = 0$. Removing the first term and working out the commutator, the following consistency condition is obtained:
\begin{equation}
    2\frac{\partial^2\Psi}{\partial x\partial y}\left(\tau + \frac{\partial^2 a}{\partial x^2}\right) + 2\left(\frac{\partial^2\Psi}{\partial x^2} - \frac{\partial^2\Psi}{\partial y^2}\right)\frac{1}{\kappa}\frac{d\kappa}{dl} + \frac{\partial^3 a}{\partial x^3} \left(\frac{\partial\Psi}{\partial y} +\frac{1}{\kappa}\frac{d\kappa}{dl}x\right) + \frac{\partial}{\partial \ell}\left(\tau - \frac{\partial^2 a}{\partial x^2} \right) = 0.
\end{equation}
The only case where it is satisfied independent of $\Psi$ is if $a(x,l) = a_0(l) + xa_1(l) - \tau x^2/2$ and $d\tau/dl = d\kappa/dl = O(\epsilon^3)$. When $\tau = 0$ and $\kappa = 1/R_0$, this consistency condition reduces to that derived in \citet{sengupta2024asymptotic}, which limited the stellarator to being a perturbed tokamak. In addition to the vertical perturbation discussed in \citet{sengupta2024asymptotic}, the condition above explicitly allows the tokamak to be perturbed in the $R$-direction; such a perturbation, if it is of the order of the minor radius, will result in an $O(\epsilon^2)$ correction to $\kappa$, which still satisfies $d\kappa/dl = O(\epsilon^3)$. Finally, when $\tau \neq 0$, the axis becomes an open helix, which corresponds to the straight stellarator limit. Similar to the tokamak, the straight stellarator can be perturbed in both the normal and binormal directions by manipulating $\kappa$ and $a_1$.

\bibliographystyle{jpp}

\bibliography{references}

\begin{thebibliography}{27}
\expandafter\ifx\csname natexlab\endcsname\relax\def\natexlab#1{#1}\fi
\def\au#1{#1} \def\ed#1{#1} \def\yr#1{#1}\def\at#1{#1}\def\jt#1{\textit{#1}}
  \def\bt#1{#1}\def\bvol#1{\textbf{#1}} \def\vol#1{#1} \def\pg#1{#1}
  \def\publ#1{#1}\def\arxiv#1{#1}\def\org#1{#1}\def\st#1{\textit{#1}}

\bibitem[Boozer(1998)]{boozer1998what}
{\sc \au{Boozer, A.~H.}} \yr{1998}  \at{What is a stellarator?}  \jt{Physics of
  Plasmas}  \bvol{5}~(5),  \pg{1647--1655}.

\bibitem[Connor {\em et~al.\/}(1978)Connor, Hastie \& Taylor]{connor1978shear}
{\sc \au{Connor, J.~W.}, \au{Hastie, R.~J.} \& \au{Taylor, J.~B.}} \yr{1978}
  \at{Shear, periodicity, and plasma ballooning modes}.  \jt{Physical Review
  Letters}  \bvol{40},  \pg{396--399}.

\bibitem[Freidberg(2014)]{freidberg2014ideal}
{\sc \au{Freidberg, J.~P.}} \yr{2014} {\em Ideal MHD\/}.  \publ{Cambridge
  University Press}.

\bibitem[Garren \& Boozer(1991{\natexlab{{\em a\/}}})]{garren1991existence}
{\sc \au{Garren, D.~A.} \& \au{Boozer, A.~H.}} \yr{1991{\natexlab{{\em a\/}}}}
  \at{Existence of quasihelically symmetric stellarators}.  \jt{Physics of
  Fluids B: Plasma Physics}  \bvol{3}~(10),  \pg{2822--2834}.

\bibitem[Garren \& Boozer(1991{\natexlab{{\em b\/}}})]{garren1991magnetic}
{\sc \au{Garren, D.~A.} \& \au{Boozer, A.~H.}} \yr{1991{\natexlab{{\em b\/}}}}
  \at{Magnetic field strength of toroidal plasma equilibria}.  \jt{Physics of
  Fluids B: Plasma Physics}  \bvol{3}~(10),  \pg{2805--2821}.

\bibitem[Helander(2014)]{helander2014theory}
{\sc \au{Helander, P.}} \yr{2014}  \at{Theory of plasma confinement in
  non-axisymmetric magnetic fields}.  \jt{Reports on Progress in Physics}
  \bvol{77}~(8),  \pg{087001}.

\bibitem[Hirshman \& Whitson(1983)]{hirshman1983steepest}
{\sc \au{Hirshman, S.~P.} \& \au{Whitson, J.~C.}} \yr{1983}
  \at{Steepest-descent moment method for three-dimensional magnetohydrodynamic
  equilibria}.  \jt{The Physics of fluids}  \bvol{26}~(12),  \pg{3553--3568}.

\bibitem[Hoelzl {\em et~al.\/}(2021)Hoelzl, Huijsmans, Pamela, B{\'{e}}coulet,
  Nardon, Artola, Nkonga, Atanasiu, Bandaru, Bhole, Bonfiglio, Cathey, Czarny,
  Dvornova, Feh{\'{e}}r, Fil, Franck, Futatani, Gruca, Guillard, Haverkort,
  Holod, Hu, Kim, Korving, Kos, Krebs, Kripner, Latu, Liu, Merkel,
  Meshcheriakov, Mitterauer, Mochalskyy, Morales, Nies, Nikulsin, Orain, Pratt,
  Ramasamy, Ramet, Reux, S\"arkim\"aki, Schwarz, Verma, Smith, Sommariva,
  Strumberger, van Vugt, Verbeek, Westerhof, Wieschollek \&
  Zielinski]{hoelzl2020the}
{\sc \au{Hoelzl, M.}, \au{Huijsmans, G.T.A.}, \au{Pamela, S.J.P.},
  \au{B{\'{e}}coulet, M.}, \au{Nardon, E.}, \au{Artola, F.J.}, \au{Nkonga, B.},
  \au{Atanasiu, C.V.}, \au{Bandaru, V.}, \au{Bhole, A.}, \au{Bonfiglio, D.},
  \au{Cathey, A.}, \au{Czarny, O.}, \au{Dvornova, A.}, \au{Feh{\'{e}}r, T.},
  \au{Fil, A.}, \au{Franck, E.}, \au{Futatani, S.}, \au{Gruca, M.},
  \au{Guillard, H.}, \au{Haverkort, J.W.}, \au{Holod, I.}, \au{Hu, D.},
  \au{Kim, S.K.}, \au{Korving, S.Q.}, \au{Kos, L.}, \au{Krebs, I.},
  \au{Kripner, L.}, \au{Latu, G.}, \au{Liu, F.}, \au{Merkel, P.},
  \au{Meshcheriakov, D.}, \au{Mitterauer, V.}, \au{Mochalskyy, S.},
  \au{Morales, J.A.}, \au{Nies, R.}, \au{Nikulsin, N.}, \au{Orain, F.},
  \au{Pratt, J.}, \au{Ramasamy, R.}, \au{Ramet, P.}, \au{Reux, C.},
  \au{S\"arkim\"aki, K.}, \au{Schwarz, N.}, \au{Verma, P.~Singh}, \au{Smith,
  S.F.}, \au{Sommariva, C.}, \au{Strumberger, E.}, \au{van Vugt, D.C.},
  \au{Verbeek, M.}, \au{Westerhof, E.}, \au{Wieschollek, F.} \& \au{Zielinski,
  J.}} \yr{2021}  \at{The {JOREK} non-linear extended {MHD} code and
  applications to large-scale instabilities and their control in magnetically
  confined fusion plasmas}.  \jt{Nuclear Fusion}  \bvol{61}~(6),  \pg{065001}.

\bibitem[Jorge {\em et~al.\/}(2020)Jorge, Sengupta \&
  Landreman]{jorge2020construction}
{\sc \au{Jorge, R.}, \au{Sengupta, W.} \& \au{Landreman, M.}} \yr{2020}
  \at{Construction of quasisymmetric stellarators using a direct coordinate
  approach}.  \jt{Nuclear Fusion}  \bvol{60}~(7),  \pg{076021}.

\bibitem[Kruger {\em et~al.\/}(1998)Kruger, Hegna \&
  Callen]{kruger1998generalized}
{\sc \au{Kruger, S.~E.}, \au{Hegna, C.~C.} \& \au{Callen, J.~D.}} \yr{1998}
  \at{{Generalized reduced magnetohydrodynamic equations}}.  \jt{Physics of
  Plasmas}  \bvol{5}~(12),  \pg{4169--4182}.

\bibitem[Landreman(2022)]{landreman2022mapping}
{\sc \au{Landreman, M.}} \yr{2022}  \at{Mapping the space of quasisymmetric
  stellarators using optimized near-axis expansion}.  \jt{Journal of Plasma
  Physics}  \bvol{88}~(6),  \pg{905880616}.

\bibitem[Landreman {\em et~al.\/}(2020--2023)Landreman, Jorge, Rodriguez \&
  Dudt]{pyQSC}
{\sc \au{Landreman, M.}, \au{Jorge, R.}, \au{Rodriguez, E.} \& \au{Dudt, D.}}
  \yr{2020--2023} py{QSC}. \url{https://github.com/landreman/pyQSC}, {S}oftware
  without accompanying journal publication.

\bibitem[Landreman \& Sengupta(2018)]{landreman2018direct}
{\sc \au{Landreman, M.} \& \au{Sengupta, W.}} \yr{2018}  \at{{Direct
  construction of optimized stellarator shapes. Part 1. Theory in cylindrical
  coordinates}}.  \jt{Journal of Plasma Physics}  \bvol{84}~(6),
  \pg{905840616}.

\bibitem[Landreman \& Sengupta(2019)]{landreman2019constructing}
{\sc \au{Landreman, Matt} \& \au{Sengupta, Wrick}} \yr{2019}  \at{Constructing
  stellarators with quasisymmetry to high order}.  \jt{Journal of Plasma
  Physics}  \bvol{85}~(6),  \pg{815850601}.

\bibitem[Landreman {\em et~al.\/}(2019)Landreman, Sengupta \&
  Plunk]{landreman2019direct}
{\sc \au{Landreman, M.}, \au{Sengupta, W.} \& \au{Plunk, G.~G.}} \yr{2019}
  \at{{Direct construction of optimized stellarator shapes. Part 2. Numerical
  quasisymmetric solutions}}.  \jt{Journal of Plasma Physics}  \bvol{85}~(1),
  \pg{905850103}.

\bibitem[Nikulsin(2021)]{nikulsin2021models}
{\sc \au{Nikulsin, Nikita}} \yr{2021}  \at{Models and methods for nonlinear
  magnetohydrodynamic simulations of stellarators}. PhD thesis, Technische
  Universit\"at M\"unchen, {A}vailable at
  \url{https://mediatum.ub.tum.de/1624218}.

\bibitem[Nikulsin {\em et~al.\/}(2021)Nikulsin, Hoelzl, Zocco, Lackner,
  G\"unter \& the JOREK~Team]{nikulsin2021testing}
{\sc \au{Nikulsin, Nikita}, \au{Hoelzl, Matthias}, \au{Zocco, Alessandro},
  \au{Lackner, Karl}, \au{G\"unter, Sibylle} \& \au{the JOREK~Team}} \yr{2021}
  \at{Testing of the new {JOREK} stellarator-capable model in the tokamak
  limit}.  \jt{Journal of Plasma Physics}  \bvol{87}~(3),  \pg{855870301}.

\bibitem[Panici {\em et~al.\/}(2023{\natexlab{{\em a\/}}})Panici, Conlin, Dudt,
  Unalmis \& Kolemen]{panici2023the}
{\sc \au{Panici, D.}, \au{Conlin, R.}, \au{Dudt, D.~W.}, \au{Unalmis, K.} \&
  \au{Kolemen, E.}} \yr{2023{\natexlab{{\em a\/}}}}  \at{The {DESC} stellarator
  code suite. {P}art 1. {Q}uick and accurate equilibria computations}.
  \jt{Journal of Plasma Physics}  \bvol{89}~(3),  \pg{955890303}.

\bibitem[Panici {\em et~al.\/}(2023{\natexlab{{\em b\/}}})Panici, Rodriguez,
  Conlin, Kim, Dudt, Unalmis \& Kolemen]{panici2023near}
{\sc \au{Panici, D.}, \au{Rodriguez, E.}, \au{Conlin, R.}, \au{Kim, P.},
  \au{Dudt, D.}, \au{Unalmis, K.} \& \au{Kolemen, E.}} \yr{2023{\natexlab{{\em
  b\/}}}} Near-axis constrained stellarator equilibria with {DESC}.  \bt{In
  {\em APS Division of Plasma Physics Meeting Abstracts\/}},  \pg{pp.
  GP11--114}.

\bibitem[Rodriguez(2022)]{rodriguez2022quasisymmetry}
{\sc \au{Rodriguez, E.}} \yr{2022}  \at{Quasisymmetry}. PhD thesis, Princeton
  University, {A}vailable at
  \url{http://arks.princeton.edu/ark:/88435/dsp01x633f4257}.

\bibitem[Rodríguez \& Bhattacharjee(2021)]{rodriguez2021solving}
{\sc \au{Rodríguez, E.} \& \au{Bhattacharjee, A.}} \yr{2021}  \at{Solving the
  problem of overdetermination of quasisymmetric equilibrium solutions by
  near-axis expansions. {II}. {C}ircular axis stellarator solutions}.
  \jt{Physics of Plasmas}  \bvol{28}~(1),  \pg{012509}.

\bibitem[Schief(2003)]{schief2003nested}
{\sc \au{Schief, W.~K.}} \yr{2003}  \at{Nested toroidal flux surfaces in
  magnetohydrostatics. {G}eneration via soliton theory}.  \jt{Journal of Plasma
  Physics}  \bvol{69}~(6),  \pg{465–484}.

\bibitem[Sengupta {\em et~al.\/}(2024{\natexlab{{\em a\/}}})Sengupta, Nikulsin,
  Gaur \& Bhattacharjee]{sengupta2024asymptotic}
{\sc \au{Sengupta, Wrick}, \au{Nikulsin, Nikita}, \au{Gaur, Rahul} \&
  \au{Bhattacharjee, Amitava}} \yr{2024{\natexlab{{\em a\/}}}}  \at{Asymptotic
  quasisymmetric high-beta three-dimensional magnetohydrodynamic equilibria
  near axisymmetry}.  \jt{Journal of Plasma Physics}  \bvol{90}~(2),
  \pg{905900209}.

\bibitem[Sengupta {\em et~al.\/}(2024{\natexlab{{\em b\/}}})Sengupta,
  Rodriguez, Jorge, Landreman \& Bhattacharjee]{sengupta2024stellarator}
{\sc \au{Sengupta, W.}, \au{Rodriguez, E.}, \au{Jorge, R.}, \au{Landreman, M.}
  \& \au{Bhattacharjee, A.}} \yr{2024{\natexlab{{\em b\/}}}}  \at{Stellarator
  equilibrium axis-expansion to all orders in distance from the axis for
  arbitrary plasma beta} (Submitted),  \arxiv{arXiv: 2402.17034}.

\bibitem[Solov'ev \& Shafranov(1970)]{solovev1970reviews}
{\sc \au{Solov'ev, L.~S.} \& \au{Shafranov, V.~D.}} \yr{1970} {\em {Reviews of
  Plasma Physics}\/}, ,  \vol{vol.~5}.  \publ{New York - London: Consultants
  Bureau}.

\bibitem[Strauss(1997)]{strauss1997reduced}
{\sc \au{Strauss, HR}} \yr{1997}  \at{Reduced {MHD} in nearly potential
  magnetic fields}.  \jt{Journal of Plasma Physics}  \bvol{57}~(1),
  \pg{83--87}.

\bibitem[Zocco {\em et~al.\/}(2020)Zocco, Helander \&
  Weitzner]{zocco2021magnetic}
{\sc \au{Zocco, A.}, \au{Helander, P.} \& \au{Weitzner, H.}} \yr{2020}
  \at{Magnetic reconnection in {3D} fusion devices: non-linear reduced
  equations and linear current-driven instabilities}.  \jt{Plasma Physics and
  Controlled Fusion}  \bvol{63}~(2),  \pg{025001}.

\end{thebibliography}

\end{document}